\documentclass[aps,pre,showpacs,reprint,groupedaddress]{revtex4-2}

\usepackage{amsmath}
\usepackage{lmodern}
\usepackage{graphicx}
\usepackage{amssymb} 
\usepackage{physics}
\usepackage{epstopdf}
\usepackage{color}
\usepackage[colorlinks,citecolor=blue,breaklinks=true]{hyperref}

\DeclareUnicodeCharacter{03B2}{\ensuremath{\beta}}

\begin{document}
\title{Mean-field equations for neural populations with $q$-Gaussian heterogeneities} 

\author{Viktoras~Pyragas and Kestutis~Pyragas}
\affiliation{Center for Physical Sciences and Technology, LT-10257 Vilnius, Lithuania} 

 

\begin{abstract}

Describing the collective dynamics of large neural populations using low-dimensional models for averaged variables has long been an attractive task in theoretical neuroscience. Recently developed reduction methods make it possible to derive such models directly from the microscopic dynamics of individual neurons. To simplify the reduction, the Cauchy distribution is usually assumed for heterogeneous network parameters.  Here we extend the reduction method for a wider class of heterogeneities defined by the $q$-Gaussian distribution. The shape of this distribution depends on the Tsallis index $q$ and gradually changes from the Cauchy distribution to the normal Gaussian  distribution as this index changes.  We derive the mean-field equations for an inhibitory network  of quadratic integrate-and-fire neurons with a $q$-Gaussian distributed excitability parameter. It is shown that the dynamic modes of the network significantly depend on the form of the distribution determined by the Tsallis index. The results obtained from the mean-field equations are confirmed by numerical simulation of the microscopic model.

\end{abstract}


\maketitle

\section{Introduction}
\label{sec:introduction}

Large systems of interacting oscillatory and/or excitable elements have been the subject of intense research in nonlinear science over the past two decades~\cite{Gupta2018,Pikovsky2015}. An important achievement in these studies belongs to  Ott and Antonsen~\cite{Ott2008}. For a large system of globally coupled heterogeneous phase oscillators (Kuramoto's model), they discovered an ansatz that allowed them to derive an exact low-dimensional system of ordinary differential equations (ODEs) to describe the averaged dynamics of a system in the thermodynamic limit. Later, Luke {\it et al}.~\cite{Luke2013} applied this approach for a  network of theta neurons. Montbri\'o {\it et al}.~\cite{Montbrio2015} considered a heterogeneous network of all-to-all pulse-coupled quadratic integrate-and-fire (QIF) neurons, which, like theta neurons, represent the canonical form of class I neurons~\cite{izhi07}. Using the Lorentzian ansatz (LA), which is different from but closely related to the Ott and Antonsen (OA) ansatz~\cite{Ott2008}, they derived a reduced system of mean-field equations for biophysically relevant macroscopic quantities, the firing rate and the mean membrane potential. 

Models for large neural populations, called neural mass models, have been developed over a long time~\cite{Wilson1973,Destexhe2009}. However, they are phenomenological in nature and do not account for the effects of synchronization between neurons. The new approach~\cite{Luke2013,Montbrio2015} makes it possible to obtain accurate reduced mean-field models directly from the microscopic dynamics of individual neurons.  
Over the past five years, these next-generation neural mass models~\cite{Coombers2019} have evolved in different directions. Reduced systems of mean-field equations were derived for excitatory neurons interacting through fast synaptic pulses of a finite width~\cite{Ratas2016}, for an inhibitory network that takes into account synaptic dynamics~\cite{Devalle2017}, in the case of constant~\cite{Pazo2016} and distributed~\cite{Ratas2018} delayed interaction, in the case of additional electrical coupling~\cite{Pietras2019,Montbrio2020}, in the presence of noise~\cite{Ratas2019,Goldobin2021a,Goldobin2021b,diVolo2022}, and for two interacting populations~\cite{Ratas2017,Segneri2020,Pyragas2021a}. Populations with heterogeneous  synaptic weights~\cite{Montbrio2015,Esnaola-Acebes2017,Bi2021,diVolo2022} and plastic synapses~\cite{Taher2020} were also considered. Reduced mean-field models are useful not only for understanding collective oscillations and other dynamic modes in large neural populations, but they can also serve as a source of simple reference systems for developing and testing various stimulation algorithms to control synchronization processes in complex networks~\cite{Pyragas2020,Pyragas2021a}.

In most publications, the reduction method is used under the assumption that the heterogeneous parameters are distributed in accordance with the Cauchy (Lorentzian) density function. The choice of such a distribution is motivated by the fact that it provides the most simple reduction. The Cauchy function has only one relevant pole in the complex plane, which leads to one equation for the complex order parameter. So far, only two recent  publications~\cite{Klinshov2021,Pyragas2021b} have considered the reduction method for non-Cauchy distributions.  In Ref.~\cite{Klinshov2021}, an approximate system of mean-field equations was obtained for an excitatory  QIF neural network with the normal Gaussian heterogeneity. To apply the theory of residues, the authors approximated the Gaussian function with a rational function by expanding its reciprocal in a truncated Taylor series. They showed that the transient dynamics and the bistability region in the parameter space change significantly compared to the Cauchy heterogeneity considered in Ref.~\cite{Montbrio2015}. In Ref.~\cite{Pyragas2021b}, we derived the mean-field equations for the same problem in the case of a bimodal heterogeneity defined by a linear combination of two Cauchy functions. We have found a wide range of dynamic modes, such as multistable equilibrium, collective oscillations and chaos, that do not exist with a unimodal distribution. Thus, the development of reduction approaches for populations with various forms of heterogeneity is an important task.

In this paper, we derive an exact system of mean-field equations for neural populations with  $q$-Gaussian heterogeneity. The $q$-Gaussian distribution is introduced in non-extensive statistical mechanics  \cite{Tsallis2009}, which   generalizes classical statistical mechanics to nonequilibrium systems. The shape of the $q$-Gaussian distribution depends on the Tsallis index $q$ and covers the Cauchy and normal Gaussian distributions as special cases. To demonstrate the reduction method with $q$-Gaussian heterogeneity,  we use an inhibitory QIF  neural network model presented in Ref.~\cite{Devalle2017} as a model of interneuronal gamma (ING) oscillations \cite{Wang1996,Whittington1995,Whittington2000,Brunel2008,Wang2010}. 

The paper is organized as follows. In Sec.~\ref{sec:q-Gaussian} we discuss $q$-Gaussian and related distributions. Section~\ref{sec:micro_model}  describes the microscopic model of the network. The derivation of the reduced mean-field equations in the thermodynamic limit is presented in Sec.~\ref{sec:mean-field}. Section~\ref{sec:bifurcation} is devoted to the bifurcation analysis of the mean-field equations. In Sec.~\ref{sec:microscopic_numerics} we present the results of numerical simulations of the microscopic model and compare them with the results obtained from the mean-field equations. A summary is given and conclusions are discussed in Sec.~\ref{sec:conclusions}.

\section{The $q$-Gaussian and related distributions}
\label{sec:q-Gaussian}

The $q$-Gaussian distribution is a probability distribution arising from the maximization of the Tsallis entropy \cite{Tsallis1988}. This entropy is the basis of non-extensive statistical mechanics \cite{Tsallis2009}, which generalizes classical statistical mechanics to nonequilibrium systems with long-range interactions and correlations. 
The $q$-Gaussian distribution  is characterized by the Tsallis index $q$ that is a measure of correlation. For $q=1$, the elements of a system are uncorrelated and the $q$-Gaussian distribution turns into normal Gaussian form. A $q$-generalization of central limit theorem was considered in Refs.~\cite{Umarov2008,Umarov2016}. Examples of successful application of the $q$-Gaussian distribution to a considerable number of various natural and artificial systems can be found in Ref.~\cite{Tsallis2012}.

The $q$-Gaussian distribution for a random variable $x$ centered at $x=0$ can be written in the form~\cite{Tsallis2009}:
\begin{equation}
p_q(x) = C_q\left[1+(q-1)\beta x^2\right]^{-1/(q-1)}. \label{q-Gaussian}
\end{equation}
Here $\beta$ is a parameter that defines the width of the distribution and $C_q$ is an appropriate normalization constant. Obviously, for $q \to 1$ the normal Gaussian distribution $p_1(x)=\sqrt{\beta/\pi}\exp(-\beta x^2)$ is recovered. In general, $q$ can be any real number less than $3$. However, here we are considering a restricted class of $q$-Gaussian distributions, assuming
\begin{equation}
q = 1+\frac{1}{n}, \label{q-n_relation}
\end{equation}
where $n=1,2,\ldots,\infty$ are natural numbers. In what  follows, we will refer to the parameter $n$  as the modified Tsallis index (MTI). The assumption \eqref{q-n_relation} turns the $q$-Gaussian distribution into a rational function. This property enables an analytic treatment of the problem presented below.  Specifically, we will consider the distribution of the excitability parameter $\eta$ of the QIF neural network in the following $q$-Gaussian form:
\begin{equation}
g_n(\eta) = C_n\left[1+\left(\frac{\eta-\bar{\eta}}{\Delta_n}\right)^2\right]^{-n}. \label{n-Gaussian}
\end{equation}
Here $\bar{\eta}$ is the center of the distribution and
\begin{equation}
\Delta_n =d\left(2^{1/n}-1\right)^{-1/2},  \label{Delta_n}
\end{equation}
where $d$ is half-width at half-maximum (HWHM) of the distribution. The form of the scale parameter $\Delta_n$ is chosen so that HWHM  does not depend on $n$, that is, it is the same for  all $n=1,2,\ldots,\infty$. This allows comparing the behavior of the QIF neural network for $q$-Gaussian distributions with different values of the parameter $n$  at fixed half-width $d$. The normalization constant in Eq.~\eqref{n-Gaussian} is
\begin{equation}
C_n = \frac{\Gamma(n)}{\sqrt{\pi} \Gamma(n-1/2) \Delta_n}, \label{C_n}
\end{equation}
where $\Gamma(\cdot)$ is the Gamma function.

\begin{figure}
\centering
	\centering\includegraphics[width=\columnwidth]{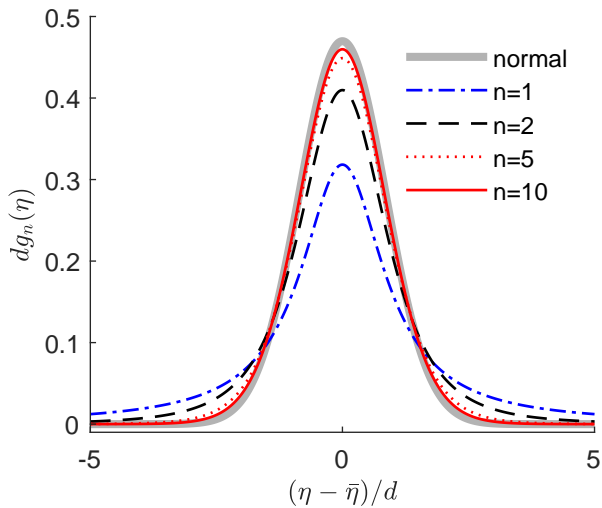}
\caption{\label{fig:q_Gauss} The $q$-Gaussian distribution \eqref{n-Gaussian} for different values of the modified Tsallis index $n$. Here $\bar{\eta}$ is the center of the distribution, and $d$ is the distribution HWHM (the same for all $n$).  For $n=1$, the $q$-Gaussian distribution coincides with the Cauchy distribution (blue dash-dotted curve), and for $n \to \infty$, with the normal Gaussian distribution (thick gray curve). 
}
\end{figure}

The distribution \eqref{n-Gaussian} is heavy tailed. For $|\eta-\bar{\eta}| \to \infty$, their tails decay by a power law $g_n(\eta) \sim |\eta-\bar{\eta}|^{-2n}$. A remarkable feature of this distribution is that for $n=1$ it coincides with the Cauchy distribution 
\begin{equation}
g_1(\eta)= \frac{1}{\pi d}\left[1+\left(\frac{\eta-\bar{\eta}}{d}\right)^2\right]^{-1}, \label{Cauchy}
\end{equation}
and as the parameter $n$ is increased to $\infty$, it gradually turns into the normal Gaussian distribution
\begin{equation}
g_\infty(\eta)=\frac{1}{d}\sqrt{\frac{\ln(2)}{\pi}} \exp\left[-\left(\frac{\eta-\bar{\eta}}{d}\right)^2\ln(2)\right]. \label{Gauss}
\end{equation}
The evolution of the distribution with increasing $n$ is shown in Fig.~\ref{fig:q_Gauss}. For $n \sim 10$, the $q$-Gaussian distribution is close to the normal Gaussian distribution.

\section{Microscopic model}
\label{sec:micro_model}

We consider a heterogeneous network of $N$ all-to-all coupled inhibitory quadratic integrate-and-fire neurons. The microscopic state of the network is given by the membrane potentials  $\{V_i\}_{i=1,\ldots,N}$, which obey the system of $N$ ordinary differential equations  of the form \cite{ermentrout10}
\begin{eqnarray}
\tau_m\dot{V}_{i} =  {V}_{i}^{2}+\eta_i-J\tau_m S(t)+I(t) \label{model}
\end{eqnarray}
with the auxiliary after-spike resetting
\begin{eqnarray}
\text{if} \; {V}_{i} \ge V_{p} \; \text{then} \; {V}_{i} \leftarrow V_{r}. \label{reset}
\end{eqnarray}
Here, $\tau_m$ is the membrane time constant, the overdot denotes the time derivative, the heterogeneous parameter of excitability $\eta_i$  is a current that specifies the behavior of each isolated neuron, $J\geq 0$ is the strength of the synaptic coupling, $S(t)$ is the synaptic activation and $I(t)$ is an external homogeneous current. The isolated neurons ($J=0$ and $I=0$) with the negative value of the parameter $\eta_i<0$ are at rest, while the neurons with the positive value of the parameter $\eta_i>0$ generate instantaneous spikes. Each time a potential $V_i$ reaches the threshold value $V_p$, it is reset to the value $V_r$, and the neuron emits an instantaneous spike which contributes to the network mean firing rate 
\begin{eqnarray}
R=\lim_{\tau_s \to 0} \frac{1}{N} \frac{1}{\tau_s} \sum_{i=1}^N \sum_k \int_{t-\tau_s}^t \delta(t'-t_i^k) d t', \label{firing_rate}
\end{eqnarray}
where $t_i^k$ is the time of the $k$th spike of the $i$th neuron, and $\delta(t)$ is the Dirac delta function. Following \cite{Devalle2017}, we take into account synaptic dynamics by a first-order differential equation for the variable $S$,
\begin{eqnarray}
\tau_d \dot{S}=-S+R, \label{mean_rate}
\end{eqnarray}
where $\tau_d$ is the synaptic time constant. The solution to this equation is a superposition of exponential post synaptic potentials emitted in the past
\begin{eqnarray}
S(t)= \frac{1}{N} \sum_{i=1}^N \sum_k H(t-t_i^k)\frac{1}{\tau_d} \exp ( \frac{t-t_i^k}{\tau_d} ), \label{synapting_activ}
\end{eqnarray}
where $H(\cdot)$ is the Heaviside step function. 

It is interesting to note that Eq.~\eqref{mean_rate} can be interpreted in terms of another physical model that takes into account the delays in the transmission of synaptic pulses between neurons instead of taking into account synaptic dynamics. This equation holds when the transmission (delay) time $\tau$ is a heterogeneous parameter distributed according to the density function $h(\tau)=\exp(-\tau/\tau_d)/\tau_d$, where $\tau_d$ is the mean time delay (see Ref.~\cite{Ratas2018} for details).

Because of the quadratic nonlinearity in Eq.~\eqref{model}, $V_j$ reaches infinity in a finite time, and this allows us to choose the threshold parameters as $V_p=-V_r=\infty$. Then the period of oscillations of an isolated neuron with $\eta_i>0$ becomes $T_i=\pi\tau_m/\sqrt{\eta_i}$.   With this choice, the QIF neuron can be transformed into a theta neuron. This choice is also crucial for the derivation of the reduced mean-field equations in the limit $N \to \infty$ \cite{Montbrio2015}.

The microscopic model described above was proposed in Ref.~\cite{Devalle2017} as a model of ING oscillations \cite{Wang1996,Whittington1995,Whittington2000,Brunel2008,Wang2010}. It was analyzed in~\cite{Devalle2017} for the case of Cauchy heterogeneity. Here, we assume that the heterogeneity of the excitability parameter $\eta_i$ is determined by the $q$-Gaussian distribution \eqref{n-Gaussian}. This distribution is characterized by three parameters:  mean $\bar{\eta}$, HWHM $d$ and MTI $n$. Using this distribution, we will derive the exact mean-field equations, and analyze how the network dynamics changes with an increase in the parameter $n$, when the form of the $q$-Gaussian distribution changes from the Cauchy to the normal Gaussian distribution.

\section{Derivation of the mean-field equations in the limit $N\to \infty$ }
\label{sec:mean-field}

In the thermodynamic limit $N\to \infty$, the infinite-dimensional system~\eqref{model} can be reduced to a low-dimensional system of ODEs using the LA method~\cite{Montbrio2015}. This method is usually applied under the assumption that the heterogeneities satisfy the Cauchy distribution \eqref{Cauchy}. In this case, the residue method allows one to reduce the network dynamics to just one equation for a complex order parameter. Recently, the LA method has been applied for more complex distributions. In Ref.~\cite{Klinshov2021}, approximate reduced equations were derived for the normal Gaussian distribution $g_\infty(\eta)$ by expanding the reciprocal $1/g_\infty(\eta)$ in a truncated Taylor series. The bimodal distribution of the excitability parameter, represented by a linear combination of two Cauchy distributions, was considered in  Ref.~\cite{Pyragas2021b}.

Here we will reproduce the LA method for the case of $q$-Gaussian heterogeneity \eqref{n-Gaussian}. The peculiarity of the $q$-Gaussian function $g_n(\eta)$ is that it  has higher-order poles on the complex plane $\eta$. As far as we know, the LA and OA reduction methods have so far been used only for the case of simple poles. Our analysis shows that these methods work for higher-order poles as well. 

In the thermodynamic limit, we characterize the population state by the density function $\rho(V|\eta,t)$, which evolves according to the continuity equation
\begin{equation}\label{cont_eq}
   \tau_m \partial_{t}\rho+\partial_{V}[(V^{2}+\eta-J \tau_m S+I)\rho]=0.
\end{equation}
According to the LA theory~\cite{Montbrio2015}, solutions of Eq.~\eqref{cont_eq} generically (independently of the initial conditions) converge to a Lorentzian-shaped function
\begin{equation}\label{rho_LA}
    \rho(V|\eta,t)=\frac{1}{\pi}\frac{x(\eta,t)}{[V-y(\eta,t)]^{2}+x^{2}(\eta,t)}
\end{equation}
with two time-dependent variables $x(\eta,t)$ and $y(\eta,t)$, which
define the half-width and the center of the voltage distribution of neurons with a given $\eta$. The ansatz~\eqref{rho_LA}  allows us  to reduce a partial differential equation~\eqref{cont_eq}  to an ODE:
\begin{equation}\label{w_eta}
    \tau_m \partial_{t}w(\eta,t) = i [ \eta+J\tau_m S(t)-w^{2}(\eta,t)+I(t)],
\end{equation}
where $w(\eta,t)=x(\eta,t)+i y(\eta,t)$ is a complex variable. The variables $x(\eta,t)$ and $y(\eta,t)$ have clear physical meanings. For a fixed $\eta$, the neurons firing rate $R(\eta,t)$ is related to the Lorentzian half-width by $R(\eta,t) = x(\eta,t)/\pi\tau_m$. This relation is obtained by estimating the probability flux $R(\eta,t) = \rho(V_p|\eta,t) \dot{V}(V_p|\eta,t)$ through the threshold $V_p=\infty$. In the thermodynamic limit, the mean firing rate \eqref{firing_rate} can be estimated as the averaged firing rate $R(\eta,t)$ over $\eta$ 
\begin{equation}\label{R_t}
    R(t)=\frac{1}{\pi \tau_m} \Re[W(t)],
\end{equation}
where
\begin{equation}\label{W_t}
    W(t)=\int^{+\infty}_{-\infty} w(\eta,t)g_n(\eta)d\eta
\end{equation}
is the averaged value of the variable $w(\eta,t)$. Here $g_n(\eta)$ is the $q$-Gaussian density function \eqref{n-Gaussian}. Averaging the variable $y(\eta, t)$ over $\eta$, we can find the mean membrane potential 
\begin{equation}\label{V_t}
    \bar{V}(t)=\Im[W(t)].
\end{equation}

Equations \eqref{w_eta}--\eqref{W_t} and \eqref{mean_rate} constitute a closed system of integro-differential equations describing the dynamics of the network in the thermodynamic limit. Further simplification can be achieved by evaluating the integral in Eq.~\eqref{W_t}. To do this, we will apply the theory of residues. Namely, the function $w(\eta,t)$ is analytically continued into a complex-valued $\eta$, and the integration contour is closed in the lower half-plane. Writing  $g_n(\eta)$ as a product of two functions
\begin{eqnarray}\label{g_n_prod}
    g_n(\eta)=C_n \Delta_n^{2n} \frac{1}{(\eta-\eta_n)^n} \frac{1}{(\eta-\eta_n^*)^n},
\end{eqnarray}
where
\begin{eqnarray}\label{eta_0}
    \eta_n=\bar{\eta} -i\Delta_n
\end{eqnarray}
and $\eta_n^*$ is the complex conjugate of $\eta_n$, we find it has two $n$ order poles, one $\eta=\eta_n$ in the lower half  plane and one $\eta=\eta_n^*$ in the higher half plane. The value of the integral \eqref{W_t} is determined by the residue at the $n$ order pole $\eta=\eta_n$ of $g_n(\eta)$ in the lower half plane:
\begin{equation}\label{W_t_1}
    W(t)=-2\pi i \frac{C_{n}\Delta^{2n}_{n}}{(n-1)!}\left[\frac{\partial^{n-1}}{\partial\eta^{n-1}}\frac{w(\eta,t)}{(\eta-\eta^{*}_{n})^{n}}\right]_{\eta=\eta_{n}}.
\end{equation}
To compute the ($n-1$)th order derivative in Eq.~\eqref{W_t_1}, we introduce $n$ time-dependent order parameters 
\begin{equation}\label{W_k_t}
W_{k}(t)=\frac{(i\Delta_{n})^{k-1}}{(k-1)!}\left[\frac{\partial^{k-1}}{\partial\eta^{k-1}}w(\eta,t)\right]_{\eta=\eta_{n}} 
\end{equation}
and $n$ coefficients
\begin{equation}\label{a_k}
a_{k}=\left[\frac{\partial^{n-k}}{\partial\eta^{n-k}}(\eta-\eta^{*}_{n})^{-n}\right]_{\eta=\eta_{n}}
\end{equation}
for $k=1,\ldots, n$. Then Eq.~\eqref{W_t_1} can be written in the form:
\begin{equation}\label{W_t_2}
W(t)=-D_{n}\sum_{k=1}^{n}\begin{pmatrix}
n-1\\
k-1
\end{pmatrix}a_k \frac{(k-1)!}{(i\Delta_{n})^{k-1}}W_{k}(t),
\end{equation}
where
\begin{equation}\label{D_n}
    D_{n}=\frac{2\pi iC_{n}\Delta^{2n}_{n}}{(n-1)!}=\frac{2i\sqrt{\pi}\Delta^{2n-1}_{n}}{\Gamma(n-\frac{1}{2})}.
\end{equation}
An explicit expression for the coefficients \eqref{a_k} is
\begin{equation}\label{a_k_1}
a_{k}=(-1)^{n}(2i\Delta_{n})^{-2n+k}\frac{(2n-k-1)!}{(n-1)!}.
\end{equation}
Finally, the function $W(t)$ can be presented by a linear combination of the order parameters $W_k(t)$,
\begin{equation}\label{W_t_3}
W(t)=\sum_{k=1}^{n}b_{k}W_{k}(t)
\end{equation}
with the coefficients
\begin{equation}\label{b_k}
b_{k}= -D_{n}\begin{pmatrix}
n-1\\
k-1
\end{pmatrix}\frac{(k-1)!}{(i\Delta_{n})^{k-1}}a_{k}.
\end{equation}
Substituting $a_k$ from  Eq.~\eqref{a_k_1} and $D_n$ from  Eq.~\eqref{D_n}, we find that these coefficients are independent of $\Delta_n$. Simplifying this equation with the Legendre duplication formula, we get the following explicit expression:
\begin{equation}\label{b_k_3}
b_{k}=\frac{\Gamma(n-\frac{k}{2})\Gamma(n-\frac{k-1}{2})}{\Gamma(n-\frac{1}{2})\Gamma(n-k+1)}, \quad k=1,\ldots,n.
\end{equation}
In numerical modeling, it is more convenient to generate these coefficients using the recurrent formula:
\begin{subequations}\label{b_k_4}
\begin{eqnarray}
b_{1}&=&1,\\
b_{k}&=&\frac{n-k+1}{n-k/2} b_{k-1}, \quad k=2,\ldots,n.
\end{eqnarray}
\end{subequations}

Now we need to derive the differential equations for the order parameters $W_k(t)$, $k=1,\ldots,n$. To do this, we differentiate  Eq.~\eqref{w_eta}  $k-1$ times by $\eta$ at the point $\eta=\eta_n$ and multiply it by the factor $[(i\Delta_n)^{k-1}/(k-1)!]$. For the left-hand side of  Eq.~\eqref{w_eta} we get
\begin{equation}\label{order_left}
\tau_m \partial_t\frac{(i\Delta_{n})^{k-1}}{(k-1)!}\left[\frac{\partial^{k-1}}{\partial\eta^{k-1}}w(\eta,t)\right]_{\eta=\eta_{n}} =\tau_m \dot{W}_k(t).
\end{equation}
On the right-hand side (RHS) of Eq.~\eqref{w_eta} there are only two terms that depend on $\eta$: $\eta$ and $w^2(\eta,t)$. Applying the above operation to the first term, we obtain:
\begin{equation}\label{order_right_1}
\frac{(i\Delta_{n})^{k-1}}{(k-1)!}\left[\frac{\partial^{k-1}}{\partial\eta^{k-1}}\eta\right]_{\eta=\eta_{n}} =\begin{cases} \eta_n &\mbox{if } k=1 \\
i\Delta_n & \mbox{if } k=2 \\
0 & \mbox{otherwise} 
\end{cases}
\end{equation}
We define the result of applying the above operation to the term $w^2(\eta,t)$ as
\begin{equation}\label{order_left_2}
Q_k(t) \equiv \frac{(i\Delta_{n})^{k-1}}{(k-1)!}\left[\frac{\partial^{k-1}}{\partial\eta^{k-1}}w^2(\eta,t)\right]_{\eta=\eta_{n}}.
\end{equation}
Performing differentiation on the RHS of this equation and using Eq.~\eqref{W_k_t}, we can express $Q_k(t)$ in terms of the order parameters as
\begin{equation}\label{Q_k}
Q_{k}(t)=\sum_{l=1}^{k}W_{k-l+1}(t)W_{l}(t), \quad k=1,\ldots,n.
\end{equation}
Summing up the above results, we obtain the following closed system of $n+1$ ODEs:
\begin{subequations}\label{w_k_ode}
\begin{eqnarray}
\tau_{m}\dot{W}_{1}&=&i[\bar{\eta}-i\Delta_{n}-J\tau_m S-W_{1}^{2}+I(t)],\label{w_k_odea}\\
\tau_{m}\dot{W}_{2}&=& -\Delta_n-i2W_1W_2,\label{w_k_odeb}\\
\tau_{m}\dot{W}_{k}&=&-iQ_{k}(t), \quad k=3,\ldots,n, \label{w_k_odec}\\
\tau_d \dot{S}&=&-S+R. \label{w_k_oded}
\end{eqnarray}
\end{subequations}
Here the expressions $Q_1=W_1^2$ and $Q_2=2W_1W_2$ are explicitly written in  Eqs.~\eqref{w_k_odea} and \eqref{w_k_odeb}. Equations.~\eqref{w_k_odea}--\eqref{w_k_odec} govern the dynamics of $n$ order parameters $[W_1(t),\ldots, W_n(t)]$.  Equation~\eqref{w_k_oded} for the synaptic variable $S(t)$ is a copy of  Eq.~\eqref{mean_rate}. We rewrote it here to complete the system of mean-field equations. The mean synaptic rate $R(t)$ is related to the order parameters according to  Eqs.~\eqref{R_t} and \eqref{W_t_3}. Recall that $n$ is the MTI of a $q$-Gaussian distribution. For $n=1$, the $q$-Gaussian distribution coincides with the Cauchy distribution, and in this case, Eqs.~\eqref{w_k_ode} are the same as the equations discussed in Ref.~\cite{Devalle2017}. In the general case, the mean-field equations~\eqref{w_k_ode} give an exact description of the macroscopic dynamics of an infinite-size network of QIF neurons satisfying the $q$-Gaussian distribution with an arbitrary modified Tsallis index $n$.

\section{Bifurcation analysis of the mean-field equations}
\label{sec:bifurcation}

We perform the bifurcation analysis of Eqs.~\eqref{w_k_ode} in the absence of external current, $I(t)=0$. For fixed $n$, these equations have five parameters: $\tau_m$, $\tau_d$, $\bar{\eta}$, $d$ and $J$. To reduce the number of parameters, we rewrite these equations in dimensionless form. Assuming the parameter $\bar{\eta}$ to be positive, we introduce the dimensionless time
\begin{equation}\label{Theta}
\vartheta= t\sqrt{\bar{\eta}}/\tau_m  
\end{equation}
and change the variables:
\begin{equation}\label{Dim_les_var}
w_{k}=\frac{W_{k}}{\sqrt{\bar{\eta}}}, \quad s=\frac{S\tau_{m}}{\sqrt{\bar{\eta}}},\quad r=\frac{R\tau_{m}}{\sqrt{\bar{\eta}}},\quad q_k=\frac{Q_k}{\bar{\eta}}.
\end{equation}
Then the system~\eqref{w_k_ode} can be presented in the form
\begin{subequations}\label{m_f_eq}
\begin{eqnarray}
w'_{1}&=&i[1-i\delta_{n}-js-w_{1}^{2}],\label{m_f_eqa}\\
w'_{2}&=& -\delta_n-i2w_1w_2,\label{m_f_eqb}\\
w'_{k}&=&-iq_{k}, \quad k=3,\ldots,n, \label{m_f_eqc}\\
\tau s'&=&-s+r, \label{m_f_eqd}
\end{eqnarray}
\end{subequations}
where the prime denotes the derivative with respect to the dimensionless time $\vartheta$ and 
\begin{subequations}\label{m_f_auxil}
\begin{eqnarray}
\delta_{n}&=&\delta(2^{1/n}-1)^{-1/2},\label{m_f_auxila}\\
q_{k}&=&\sum_{l=1}^{k}w_{k-l+1}w_{l},\label{m_f_auxilb}\\
r&=&\frac{1}{\pi}\Re\sum_{l=1}^{n}b_{l}w_{l}.\label{m_f_auxilc}
\end{eqnarray}
\end{subequations}
The coefficients $b_l$ are determined by  Eq.~\eqref{b_k_3} or by the recurrent formula~\eqref{b_k_4}. In the new variables, the mean membrane potential $v=\bar{V}/\sqrt{\bar{\eta}}$  becomes
\begin{equation}\label{v_t}
    v=\Im\sum_{l=1}^{n}b_{l}w_{l}.
\end{equation}
The advantage of the system \eqref{m_f_eq} over the system \eqref{w_k_ode} is that it depends on only three parameters:
\begin{equation}
j=J/\sqrt{\bar{\eta}},\quad \tau=\sqrt{\bar{\eta}}\tau_{d}/\tau_{m},\quad \delta=d/\bar{\eta}.
\end{equation}
The parameter $j$ is the new normalized coupling strength, $\tau$ is proportional to the ratio of the synaptic time constant $\tau_d$ to the most-likely period $\bar{T}=\pi\tau_m/\sqrt{\bar{\eta}}$ of the neurons, and $\delta$ is  the ratio of the half-width $d$ to the center $\bar{\eta}$ of the $q$-Gaussian distribution. Next, we will analyze how the solutions of  Eqs.~\eqref{m_f_eq} depend on these parameters.

We start the analysis of the mean-field equation by determining the equilibrium points and their stability. We denote the equilibrium solution of Eqs.~\eqref{m_f_eq} by a tilde: $(\tilde{w}_1,\ldots, \tilde{w}_n,\tilde{s})$.  It is obtained by equating the RHS of these equations to zero. In the general case, this problem requires solving a system of polynomial equations. However, if we are interested in the dependence of the equilibrium points on the coupling strength $j$, we do not need to solve the polynomial equations. Defining an independent parameter as $p=j\tilde{s}$, we can find the dependence of the equilibrium values of $\tilde{w}_{k}$ on this parameter:
\begin{subequations}\label{equilibr1}
\begin{eqnarray}
\tilde{w}_{1}(p)&=&\sqrt{1-i\delta_{n}-p},\\
\tilde{w}_{2}(p)&=& i\delta_{n}/2w_{1}(p),\\
\tilde{w}_{k}(p)&=&\frac{-1}{\tilde{w}_{1}(p)}\sum^{k-1}_{l=2}\tilde{w}_{k-l+1}(p)\tilde{w}_{l}(p),\: k=3,..,n
\end{eqnarray}
\end{subequations}
Taking into account that the equilibrium values of $s$ and $r$ coincide, $\tilde{s}=\tilde{r}$, we can parametrically establish the dependence of the equilibrium spiking rate on the coupling strength:  
\begin{subequations}\label{equilibr2}
\begin{eqnarray}
\tilde{r}(p)&=&\frac{1}{\pi}\Re\sum^{n}_{l=1}b_{l}\tilde{w}_{l}(p), \label{equilibr2a}\\
j(p)&=&p/\tilde{r}(p) \label{equilibr2b}.
\end{eqnarray}
\end{subequations}

\begin{figure}
\centering\includegraphics[width=\columnwidth]{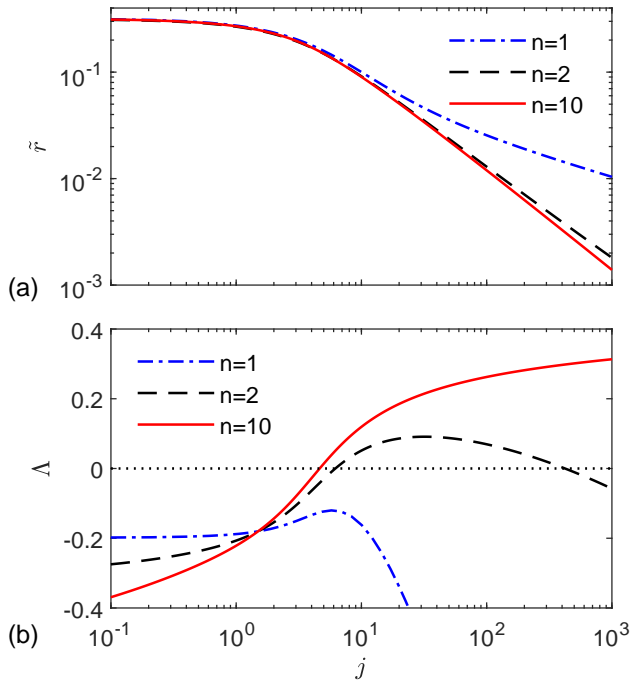}
\caption{\label{linear_stab} Linear stability of the equilibrium state of the  mean-field equation~\eqref{m_f_eq} of the QIF neuron network: (a) the equilibrium spiking rate $\tilde{r}$ and (b) the maximum real part $\Lambda$ of the eigenvalues of the linearized system \eqref{lin_m_f1} as functions of the coupling strength $j$ for fixed parameters $\tau=2$ and $\delta=0.2$ and different values of the modified Tsallis index $n$.
}
\end{figure}

Figure~\ref{linear_stab}(a) shows this dependence at fixed parameters $\tau=2$ and $\delta=0.2$ and different values of the MTI. The network has a single equilibrium point for any $j$ and $n$. For $n\geq 2$, the characteristics  $\tilde{r}(j)$   are practically independent of $n$,  while in the case of the Cauchy distribution ($n=1$) this characteristic is significantly different. Here, the spiking rate $\tilde{r}$ decreases much more slowly with increasing $j$. This is due to the fact that the Cauchy distribution has especially heavy tails, and even with a large inhibitory coupling strength $j$, a significant number of neurons remain in the right heavy tail, where they are active. 

Next we analyze how the stability of the equilibrium state depends on $j$. We introduce small deviations $\delta w_k = w_k-\tilde{w}_k$ and $\delta s=s-\tilde{s}$ from the equilibrium state, and linearize the system~\eqref{m_f_eq} with respect to these deviations. To simplify expressions, we define $n$-dimensional vectors $\delta {\bf{w}}=(\delta w_1,\ldots, \delta w_n)^T$, ${\bf{b}}=(b_1,\ldots, b_n)^T$ and ${\bf{e}}=(1, 0,\ldots, 0)^T$, where the superscript $T$ denotes the transpose operation, and write the linearized system as:    
\begin{subequations}\label{lin_m_f}
\begin{eqnarray}
\delta{{\bf{w}'}}&=&-2i{\bf{W}}\delta {\bf{w}}-ij{\bf{e}}\delta s, \label{lin_m_fa}\\
\tau\delta{s'}&=&-\delta s+{\bf{b}}^{T}\Re(\delta {\bf{w}})/\pi.\label{lin_m_fb}
\end{eqnarray}
\end{subequations}
Here
\begin{eqnarray} \label{W_matr}
{\bf{W}}=\begin{pmatrix} 
\tilde{w}_{1} & 0 & 0& \cdots & 0\\
\tilde{w}_{2} & \tilde{w}_{1}& 0 & \cdots& 0\\
\tilde{w}_{3}& \tilde{w}_{2} & \tilde{w}_{1}&\cdots& 0\\
\vdots& \vdots &\vdots &\vdots &\vdots\\
\tilde{w}_{n}& \tilde{w}_{n-1}&\tilde{w}_{n-2}&\cdots&\tilde{w}_{1}
\end{pmatrix}
\end{eqnarray}
is an $n\times n$ lower triangle matrix. Having written the complex vector $\bf{w}$ and the complex matrix $\bf{W}$ as
\begin{subequations}\label{w_W}
\begin{eqnarray}
\delta {\bf{w}}&=&\delta {\bf{u}}+i\delta {\bf{v}},\label{w_Wa}\\
{\bf{W}}&=&{\bf{U}}+i{\bf{V}}, \label{w_Wb}
\end{eqnarray}
\end{subequations}
Eqs.~\eqref{lin_m_f} can finally be presented as a linear system of $2n+1$ real ODEs:
\begin{subequations}\label{lin_m_f1}
\begin{eqnarray}
\delta{{\bf{u}}'}&=&2({\bf{V}}\delta {\bf{u}}+{\bf{U}}\delta {\bf{v}}), \label{lin_m_f1a}\\
\delta{{\bf{v}}'}&=&2({-\bf{U}}\delta {\bf{u}}+{\bf{V}}\delta {\bf{v}})-j{\bf{e}}\delta s, \label{lin_m_f1b}\\
\delta{s'}&=&(-\delta s+{\bf{b}}^{T}\delta {\bf{u}}/\pi)/\tau. \label{lin_m_f1c}
\end{eqnarray}
\end{subequations}
Solving the eigenvalue problem for this system  at a given value of the parameter $p$, we obtain $2n+1$ eigenvalues $\lambda_m(p)$, $m=1,\ldots,2n+1$. Equation~\eqref{equilibr2b} gives the dependence of the coupling strength on the parameter $j(p)$, and thus we have a parametric dependence of $\lambda_m$ on $j$. The equilibrium state is stable if the real parts of all eigenvalues are negative.

Figure~\ref{linear_stab}(b) shows the dependence of the maximum real part 
\begin{equation}\label{Lambda}
\Lambda=\max_m [\Re(\lambda_m)]  
\end{equation}
of the eigenvalues on the coupling strength $j$ for the same values of $\tau$, $\delta$ and $n$ as in Fig.~\ref{linear_stab}(a). For $n=1$, the value of $\Lambda$ is negative at any $j$. Thus, the Cauchy distribution provides the stable equilibrium state at any coupling strength $j$, and for given $\tau$ and $\delta$ the network cannot produce macroscopic oscillations. For $n\geq 2$, the stability properties  of the  equilibrium state [characteristic $\Lambda(j)$] essentially depend on $n$, although the characteristic $\tilde{r}(j)$ is almost independent of $n$. The values of $j$ where $\Lambda(j)=0$ represent the Hopf bifurcation points. In the regions where $\Lambda(j)>0$, the network exhibits macroscopic limit cycle oscillations. The interval of $j$ where oscillations take place enlarges with the increase of $n$. The effect of improving the oscillatory properties is due to the fact that an increase in $n$ makes the tails of the $q$-Gaussian distribution less heavy, so that more neurons are concentrated in the center of the distribution, and these neurons are responsible for the appearance of collective oscillations. 

\begin{figure}
\centering
	\centering\includegraphics[width=\columnwidth]{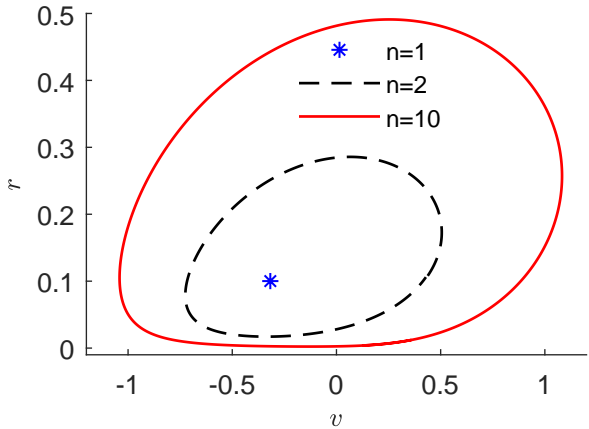}
\caption{\label{fig:fp_v_r} Attractors of the mean-field equations~\eqref{m_f_eq} depending on the modified Tsallis index: fixed point at $n=1$, small limit cycle at $n=2$, and large limit cycle at $n=10$. The variables $r$ and $v$ represent the dimensionless spiking rate and the mean membrane potential, respectively. The parameter values are $\tau=2$, $\delta=0.2$, and $j=10$.
}
\end{figure}

Figure~\ref{fig:fp_v_r} shows asymptotic solutions of the nonlinear system of  mean-field equations~\eqref{m_f_eq} in the plane of the variables $r$ and $v$ for different values of the MTI $n$. The parameters $\tau=2$ and $\delta=0.2$ are the same as in Fig.~\ref{linear_stab}, and the coupling strength is $j=10$. For $n=1$, the equilibrium state is stable ($\Lambda<0$), and the solutions of the nonlinear system converge to a fixed point indicated by a blue asterisk. For $n=2$ and $10$ the equilibrium state is unstable ($\Lambda>0$) and the solutions converge to limit cycles shown by black dashed and red solid curves, respectively. The size of the limit cycle increases with increasing $n$. For $n\to \infty$ the size of the limit cycle saturates to the size obtained from the microscopic model with a normal Gaussian distribution (see Figs.~\ref{fig:QIFvsMF} and \ref{fig:QIFvsMF_Gauss} below). Note that the effect of increasing $n$ (making the distribution less heavy-tailed) to the occurrence  of collective oscillations is somewhat similar to the effect of noise in Refs.~\cite{Ratas2019,Goldobin2021a,Goldobin2021b,diVolo2022}, where it was shown that non-oscillating networks with Cauchy-distributed parameters can oscillate in the presence of noise.

\begin{figure*}
\centering\includegraphics[width=\textwidth]{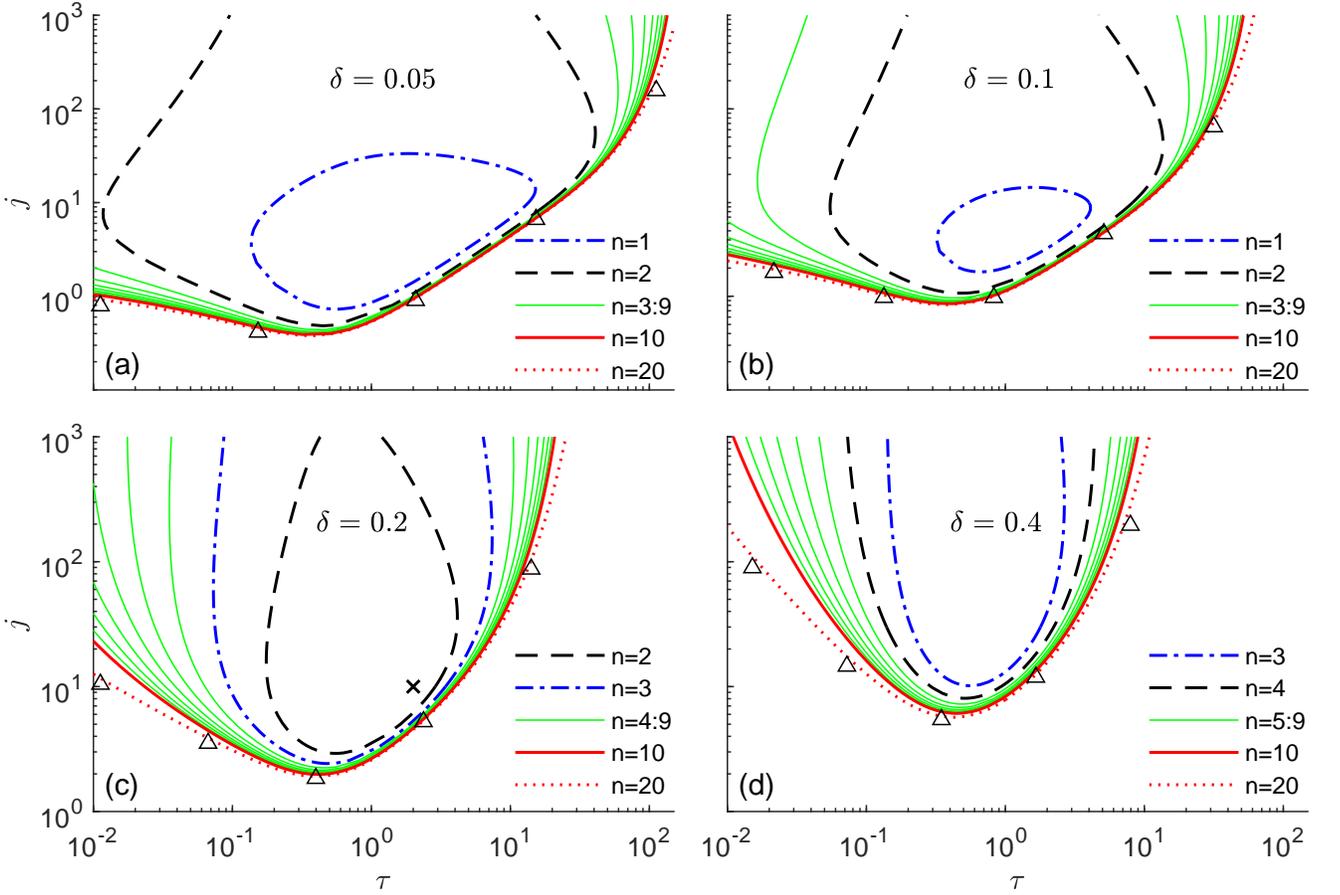}
\caption{\label{two_par_bif} Two-parameter bifurcation diagrams of the mean-field equations~\eqref{m_f_eq} in the plane of parameters $(\tau, j)$ for various fixed values of the parameter $\delta$: (a) $0.05$, (b) $0.1$, (c) $0.2$ and (d) $0.4$.  Lines of different styles indicate Hopf bifurcation curves with different values of the modified Tsallis index $n$. Triangles represent estimates of the bifurcations' loci obtained by direct numerical simulations of the microscopic model \eqref{model} with $N=5\times 10^4$ QIF neurons distributed according to the normal Gaussian density function (see Sec.~\ref{sec:microscopic_numerics} for details). The cross in (c) indicates the parameter values used in numerical simulations of Figs.~\ref{fig:fp_v_r}, \ref{fig:QIFvsMF}, and \ref{fig:QIFvsMF_Gauss}.} 
\end{figure*}

Relatively simple mean-field equations~\eqref{m_f_eq} make it possible to carry out a two-parameter bifurcation analysis of the system even for sufficiently large values of the MTI $n$. Figure~\ref{two_par_bif} shows the bifurcations diagrams  in the parameter plane $(\tau, j)$. They were built using the MATCONT package~\cite{matcont}. The diagrams are presented in four panels for four different fixed values of the parameter $\delta$: (a) $0.05$, (b) $0.1$, (c) $0.2$, and (d) $0.4$.  Lines of different styles indicate Hopf bifurcation curves with different values of $n$. They divide the $(\tau, j)$ plane into regions with a stable equilibrium state and stable limit cycle oscillations. For all $\delta$, the region of the limit cycle oscillations increases with increasing $n$ and reaches a maximum at $n\to\infty$, when the $q$-Gaussian distribution goes over to the normal distribution. 

For fixed $n$ and increasing $\delta$, the region of the limit cycle oscillations narrows and disappears at some threshold value $\delta=\delta_n$. For example, the evolution of this region for the Cauchy distribution ($n=1$)  is seen in Figs. \ref{two_par_bif}(a) and \ref{two_par_bif}(b) from the change in the Hopf bifurcation curve, indicated by the blue dash-dotted line. Closed bifurcation curves mean that collective oscillations occur only in a limited range of parameters. They cannot occur if the synapse is too slow (large $\tau$) or too fast (small $\tau$). Oscillations of the limit cycle are also impossible if the coupling strength $j$ is too small or too large. At $\delta=0.2$ [Fig.\ref{two_par_bif}(c)], collective oscillations cannot appear for any $\tau$ and $j$: At $n=1$  the region of limit cycle oscillations disappears. This result is consistent with the analytical conclusions of Ref.~\cite{Devalle2017} where it was shown that the limit cycle oscillations cannot exist in a system with Cauchy heterogeneity if $\delta>\delta_1\approx 0.14$. However, a $q$-Gaussian distribution with MTI $n=2$ preserves the limit cycle oscillations beyond this threshold. At $n=2$, the oscillations disappear at higher values of $\delta$, when  $\delta>\delta_2\approx 0.36$. With a further increase in $\delta$, the evolution of bifurcation curves with a higher MTI is similar: The oscillation regions narrow and disappear one after another at $n$ equal to $3$, $4$, etc. Surviving regions with large $n$ move towards higher values of the coupling strength $j$.  

Summarizing the above bifurcation analysis, we emphasize that the oscillatory properties of the network significantly depend on the shape of the distribution determined by the modified Tsallis index $n$. For a fixed ratio $\delta=d/\bar{\eta}$, the domain of the limit cycle oscillations in the parameter space increases with  an increase in  $n$. Among the family of distributions covered by a $q$-Gaussian function, the Cauchy distribution ($n=1$) is less effective for generating collective synchronized oscillations, while the normal distribution ($n=\infty$) is the most efficient.

\section{Comparison of solutions of the mean-field equations and the microscopic model}
\label{sec:microscopic_numerics}

\begin{figure}
\centering
	\centering\includegraphics[width=\columnwidth]{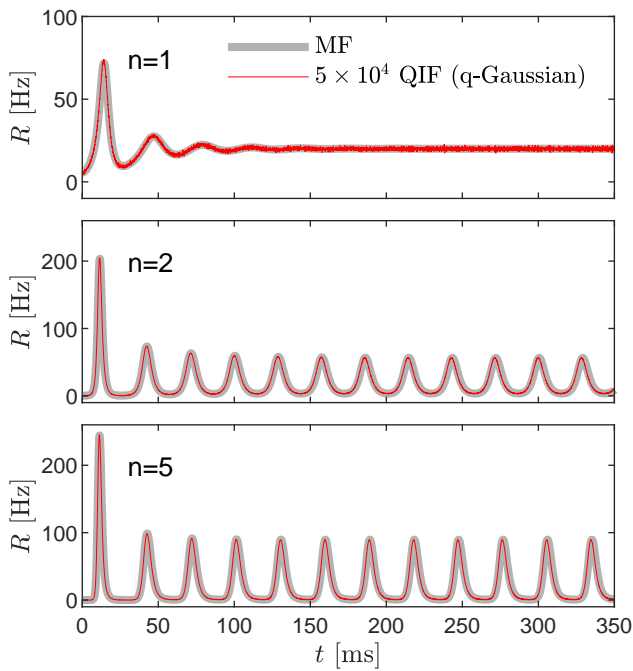}
\caption{\label{fig:QIFvsMF} Comparison of the dynamics of the spiking rate of the microscopic model \eqref{theta_j} (thin red curves) and mean-field equations~\eqref{w_k_ode} (thick gray curves) with $q$-Gaussian heterogeneity at different values of the modified Tsallis index $n$. The parameter values are $\tau_m=10$ ms, $\tau_d=10$ ms, $\bar{\eta}=4$, $d=0.8$ and $J=20$.
}
\end{figure}

\begin{figure}
\centering\includegraphics[width=\columnwidth]{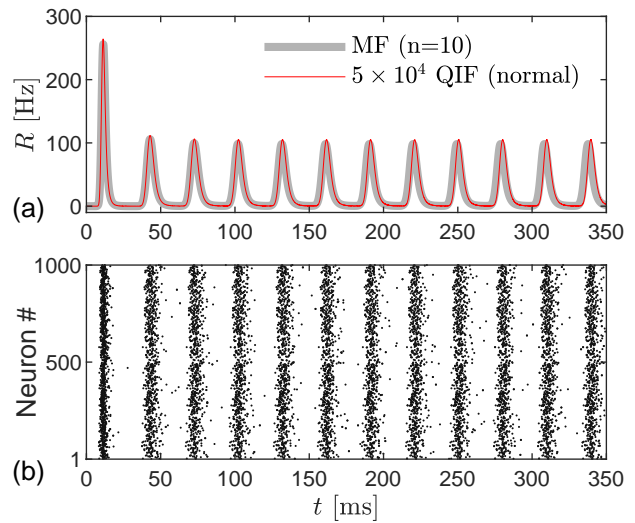}
\caption{\label{fig:QIFvsMF_Gauss} Dynamics of a population of normally distributed $5\times 10^4$ neurons and its approximation by mean-field equations.
The values of the parameters are the same as in Fig.~\ref{fig:QIFvsMF}. (a)~Firing rate obtained from the microscopic model \eqref{theta_j} (thin red curve) with the normal Gaussian distribution Eq.~\eqref{Gauss}	 and from the mean-field equations~\eqref{w_k_ode} (thick gray curve) with the $q$-Gaussian distribution \eqref{n-Gaussian} at $n=10$. (b)~Raster plot of $1000$ randomly selected neurons. The dots show the spike moments for each neuron, where the vertical axis indicates neuron numbers.
}
\end{figure}

The reduced mean-field  equations~\eqref{w_k_ode} are exact in the limit of an infinite-size network. Here we verify how well they describe the dynamics of  
finite-size networks described by the microscopic model \eqref{model} and \eqref{mean_rate}.  Numerical simulation of these equations is more convenient after changing the variables
\begin{equation}
V_i = \tan(\theta_i/2)\label{eq_transf_tet}
\end{equation}
that turn QIF neurons into theta neurons. Such a transformation of variables avoids the problem of resetting the membrane potential $V_i$ of the QIF neuron from  $+ \infty $ to $-\infty $ at the moments of firing. At these moments, the phase $\theta_i$ of the theta neuron simply crosses the value of $\theta_i=\pi$. For theta neurons,  Eqs.~\eqref{model} and \eqref{mean_rate} are transformed into
\begin{subequations} \label{theta_j}
\begin{eqnarray}
\tau_m \dot{\theta}_{i}&=& 1-\cos (\theta_{i})\nonumber\\
&+&\left[1+\cos (\theta_{i})\right]\left[\eta_{i} -J\tau_m S+I(t) \right], \label{theta_ja}\\ 
\tau_d \dot{S}&=& -S+R, \label{theta_jb}
\end{eqnarray}
\end{subequations}
where $R$ is the mean firing rate defined by  Eq.~\eqref{firing_rate}. For the numerical implementation of this equation, we set $\tau_s =  10^{-2} \tau_m$. To obtain a smoother time series, the firing rate plotted in Figs.~\ref{fig:QIFvsMF} and \ref{fig:QIFvsMF_Gauss} was computed according to Eq.~\eqref{firing_rate} with $\tau_s = 3\times 10^{-2} \tau_m $. The parameter values $\{\eta_i\}_{i=1,\ldots,N}$ satisfying the $q$-Gaussian distribution $g_n(\eta)$ Eq.~\eqref{n-Gaussian} were deterministically generated using $\eta_i=\bar{\eta}+G_n^{-1}(\xi_i)$, where $\xi_{i=1,\ldots,N}=i/(N+1)$  are the numbers uniformly distributed in the unity interval $\xi \in (0,1)$ and $G_n^{-1}$ denotes the inverse of the cumulative $q$-Gaussian distribution function $G_n(\eta)$ with $\bar{\eta}=0$, i.e., $G_n(\eta)=C_n\int_{-\infty}^\eta [1+(\eta'/\Delta_n)^2]^{-n} d \eta'$. Equations~\eqref{theta_j} were integrated by the Euler method with a time step of $d t = 10^{-4} \tau_m$. 

In Fig.~\ref{fig:QIFvsMF} we compare the solutions of the microscopic model \eqref{theta_j} and the mean-field equations~\eqref{w_k_ode} for three different MTI values: $n=1,2$ and $5$. The parameter values $\tau_m=10$ ms, $\tau_d=10$ ms, $\bar{\eta}=4$, $d=0.8$, and $J=20$ are chosen such that they give the same values of  the dimensionless parameters $\tau=2$, $\delta=0.2$ and $j=10$, which  were used in Fig.~\ref{fig:fp_v_r}. To match the initial conditions of the microscopic model and the mean-field equations  at $t=0$, we proceeded as follows. We turned on the constant inhibitory current $I(t)=-4$ at $t=-200$ ms and, using arbitrary initial conditions, integrated both systems in the interval $t \in [-200, 0]$ ms. Due to the inhibitory current,  both systems reached the same stable equilibrium state at the end of the interval. Then, at $t=0$,  we turned off the inhibitory current $I(t)=0$, and for $t>0$, we calculated the dynamics of the spiking rate $R(t)$ for both systems. Our results show that the dynamics  of the spiking rate calculated using the microscopic model with  $N=5\times 10^4$ neurons is in good agreement with the dynamics obtained from the mean-field equations. For a given $\delta=0.2$, the Cauchy distribution ($n=1$) cannot provide collective oscillations in the network and the system relaxes to a stable incoherent state with a constant spiking rate. For the same parameter values, $q$-Gaussian distributions with higher values of $n=2$ and $5$ lead to synchronized limit cycle oscillations.   

In Fig.~\ref{fig:QIFvsMF_Gauss}(a) we compare the solutions of the  microscopic model  \eqref{theta_j} and the mean-field equations~\eqref{w_k_ode}, with the excitability parameter of the microscopic model described by a normal Gaussian distribution, and the mean-field model corresponding to a $q$-Gaussian distribution with $n=10$. Good agreement between these solutions indicates that the dynamics of a large-scale network of normally distributed QIF neurons can be well approximated by a low-dimensional system of mean-field equations represented by a $q$-Gaussian distribution with MTI $n \sim 10$. The raster plot shown in Fig.~\ref{fig:QIFvsMF_Gauss}(b) demonstrates network dynamics at the microscopic level. Here, dots denote spike moments of $10^3$ of randomly selected neurons out of $5\times 10^4$ of the total number of neurons.

In addition, we tested the possibility of mean-field equations for predicting the onset of oscillations in a finite size network with a normal Gaussian distribution of the excitability parameter. The results obtained from  the microscopic model \eqref{model} with $N=5\times 10^4$ normally distributed neurons are shown in Fig.~\ref{two_par_bif} by triangles. They denote  numerically estimated loci of Hopf bifurcations for several values of the parameters $\tau$ and $\delta$. For  $\delta \leq 0.1$, these symbols are close to the Hopf bifurcation curves of the mean-field equations with the MTI $n=10$ [Figs.~\ref{two_par_bif}(a) and \ref{two_par_bif}(b)]. Thus, the mean-field equations with only ten order parameters, approximate well the boundaries of oscillations of a large network with a normal Gaussian heterogeneity. For $\delta> 0.1$, more order parameters ($n=20$) may be required to approximate well the Hopf bifurcation loci [Figs.~\ref{two_par_bif}(c) and \ref{two_par_bif}(d)].

\section{Discussion}
\label{sec:conclusions}

We  derived a reduced system of mean-field equations for an inhibitory QIF neural network with a heterogeneous excitability parameter $\eta$ distributed according to a $q$-Gaussian density function. We considered a restricted class of $q$-Gaussian distributions with the Tsallis index $q=1+1/n$, where $n$ is any natural number, here called the modified Tsallis index. With this restriction, the $q$-Gaussian distribution $g_n(\eta)$ is a rational function, which allows us to apply the theory of residues. The peculiarity of our case is that the $q$-Gaussian function $g_n(\eta)$ has poles of order $n$ on the complex plane $\eta$. Until now, the LA reduction method had only been used for simple poles. Our analysis showed  that this method also works for higher order poles and leads to relatively simple mean-field equations. It is noteworthy that for $n=1$ the $q$-Gaussian distribution coincides with the Cauchy distribution, and as $n$ is increased to infinity, it gradually turns into the normal Gaussian distribution. Comparing the results for a family of $q$-Gaussian distributions with different MTIs $n$, we assume that the center $\bar{\eta}$ and half-width $d$ of the distribution are the same for all $n$.

The method presented here for deriving the mean field equations can be applied to various network models with $q$-Gaussian distributed parameters. In this paper, we demonstrated this on a specific ING oscillation model, which was previously considered for the case of Cauchy heterogeneity~\cite{Devalle2017}. We performed the bifurcations analysis  of this system depending on the MTI $n$ and other parameters of the model. We found that the collective oscillation regions in the parameter space expand with increasing $n$, as the $q$-Gaussian distribution changes from the Cauchy distribution to the normal distribution. In particular, for larger $n$, oscillations occur in wider ranges of the synaptic time $\tau_d$ and the coupling strength $J$.  Previously it was found that collective oscillations cannot occur in a system with a Cauchy distribution if the parameter $\delta=d/\bar{\eta}$ exceeds a certain threshold $\delta_1 \approx 0.14$~\cite{Devalle2017}. Here we have shown that $q$-Gaussian distributions with $n>1$ provide oscillations beyond this threshold. As a general conclusion, we note that the oscillatory properties of the network significantly depend on the shape of the distribution determined by the MTI $n$. Among the distributions covered by the $q$-Gaussian family, the normal distribution is most efficient for generating collective oscillations, while the Cauchy distribution is less efficient. 

The mean-field equations derived here are exact in the thermodynamic limit $N \to \infty $ for $q$-Gaussian distributions with any modified Tsallis index $n$. The $n$ index determines the number of ODEs in the reduced mean-field model. To verify how well the mean-field equations describe the dynamics of finite-size networks, we numerically simulated the equations of the microscopic model. Modeling networks with $ N = 5 \times 10^4$ inhibitory $q$-Gaussian-distributed QIF neurons at various values of $n$ yielded results that are in good agreement with the results obtained from the mean-field equations. 

In addition, we simulated networks with normally distributed QIF neurons and tested how well the results obtained can be approximated by mean-field equations with finite $n$. This simulation showed that the mean-field equations valid for a $q$-Gaussian distribution  with $n \sim 10$ approximate well the dynamics of large normally distributed neural populations. Note that Ref.~\cite{Klinshov2021} considered an excitatory QIF neural network with a normal distribution approximated by a rational function having simple distinct poles. In that approximation, the values of the poles were determined numerically and the resulting system of mean-field equations had coefficients that also required numerical calculation. The advantage of approximating the normal Gaussian distribution by a $q$-Gaussian function is that it has one explicit higher-order pole, and the system of mean-field equations obtained here has a simple form with explicitly given coefficients.

\begin{acknowledgements}
This work was supported by the Research Council of Lithuania through Grant No. S-MIP-21-2.
\end{acknowledgements}

\bibliography{QIF_q_Gaussian}

\end{document}